\documentclass[manuscript]{aastex}
\usepackage{lscape}

\shorttitle{Spectroscopy of SMC Red Giants. II.}
\shortauthors{Parisi et al.}

\begin{document}

\title{\ion{Ca}{2} Triplet Spectroscopy of Small Magellanic Cloud Red Giants. II.
Abundances for a Sample of Field Stars. }

\author{M.C. Parisi}
\affil{Observatorio Astron\'omico, Universidad Nacional de C\'ordoba}
\affil{Laprida 854, C\'ordoba, CP 5000, Argentina.}
\email{celeste@oac.uncor.edu}

\author{D. Geisler}
\affil{Departamento de Astronom{\'\i}a, Universidad de Concepci\'on}
\affil{Casilla 160-C, Concepci\'on, Chile}
\email{dgeisler@astro-udec.cl}

\author{A.J. Grocholski}
\affil{Space Telescope Science Institute}
\affil{3700 San Martin Dr., Baltimore, MD 21218, USA.}
\email{aarong@stsci.edu}

\author{J.J. Clari\'a}
\affil{Observatorio Astron\'omico, Universidad Nacional de C\'ordoba}
\affil{Laprida 854, C\'ordoba, CP 5000, Argentina.}
\email{claria@oac.uncor.edu}

\and

\author{A. Sarajedini}
\affil{Department of Astronomy, University of Florida}
\affil{PO Box 112055, Gainesville, FL 32611, USA.}
\email{ata@astro.ufl.edu}

\begin{abstract}

We have obtained metallicities of $\sim$ 360 red giant stars distributed in 15 
Small Magellanic Cloud (SMC) fields from near-infrared spectra covering 
the \ion{Ca}{2} triplet lines using the VLT + FORS2. The errors of the 
derived [Fe/H] values range from 0.09 to 0.35 dex per star, with a mean of 0.17 
dex. The metallicity distribution of the whole sample shows a mean value 
of [Fe/H] = -1.00 $\pm$ 0.02, with a dispersion of 0.32 $\pm$ 0.01, 
in agreement with global mean  [Fe/H] values found in previous studies. 
We find no evidence of a metallicity gradient in the SMC. In fact, on 
analysing the metallicity distribution of each field, we derived mean values 
of [Fe/H] = -0.99 $\pm$ 0.08 and  [Fe/H] = -1.02 $\pm$ 0.07 for fields located 
closer and farther than 4$^{\circ}$ from the center of the galaxy, respectively. 
In addition, there is a clear tendency for the field stars to be more metal-poor than 
the corresponding cluster they surround, independent of their positions in 
the galaxy and of the clusters' age.
We argue that this most likely stems from the field stars being somewhat older
and therefore somewhat more metal-poor than most of our clusters.

\end{abstract}


\section{Introduction}

Local Group galaxies have long been recognized as being excellent laboratories 
to understand the star formation and the chemical enrichment histories of dwarf 
galaxies. In particular, the Small Magellanic Cloud (SMC) and its companion, the
Large Magellanic Cloud (LMC), are close enough to resolve their oldest individual 
stars, thus allowing a detailed determination of the full range of ages as well
as metallicities. This 
permits a better understanding of the formation and evolution of this kind of galaxy. 
Unfortunately, the SMC has not been studied as thoroughly as the LMC. Our knowledge 
about the chemical evolution history of the SMC mainly comes from the study of its 
cluster system. Da Costa \& Hatzidimitriou (1998) and Piatti et al. (2001, 2005, 2007a,b,c),
among others, have tried to derive  the Age-Metallicity Relation (AMR) of this 
galaxy on the basis of \ion{Ca}{2} triplet spectroscopic and photometric studies, 
respectively. Recently, Parisi et al. (2009) (hereafter Paper I), applied the \ion{Ca}{2} 
triplet method (Cole et al. 2004; Grocholski et al. 2006) to 15 SMC clusters. The AMR derived  
in Paper I shows evidence for 3 phases: a very early ($>11$ Gyr) phase in which the 
metallicity reached [Fe/H] $\sim$ -1.2; a long intermediate phase from $\sim 10-3$ 
Gyr in which the metallicity only slightly increased although a number of clusters 
formed, and a final phase from 3-1 Gyr ago 
in which the rate of enrichment was remarkably faster. They find good overall 
agreement with the model of Pagel \& Tautvai\v{s}ien\.{e} (1998)  which assumes a burst of star formation at 4 Gyr.

There are just a few studies in which the chemical enrichment history of the SMC is 
analyzed from field stars. Dolphin et al. (2001), based on $VI$ photometry of a field 
in the outer SMC, found a metallicity of [Fe/H] = -1.3 $\pm$ 0.3 for the oldest stars, which 
increased up to [Fe/H] = -0.7 $\pm$ 0.2 by 1-2 Gyr ago. Harris \& Zaritsky (2004)  
derived the chemical enrichment history in the central area of the SMC based on $UBVI$ 
photometry from their Magellanic Cloud Photometric Survey. They found that the stars 
formed until $\sim$ 3 Gyr ago have a mean abundance [Fe/H] $\sim$ -1 rising monotonically to 
a present value of [Fe/H] $\sim$ -0.4. Using CaT spectroscopy of $\sim$ 350 red giant 
branch stars in 13 SMC fields, Carrera et al. (2008) found that in the innermost fields the average 
metallicity is [Fe/H] $\sim$ -1.0. However, this value decreases in the  
outermost 
regions, suggesting a metallicity gradient.
They also showed that this metallicity gradient is related to an age gradient 
in the sense that the stars concentrated in the central regions are generally
younger. Carrera et al. (2008)'s 
study supports the results of Piatti et al. (2007a,b) who also came to a similar conclusion. However, no evidence of such a gradient was found in Paper I, which covered a much
wider range in galactocentric distance. 
The recent work of Cioni (2009), using the C/M ratio of asymptotic giant
branch (AGB) stars to derive metallicity, also supports a negligible gradient.

It is interesting to compare the metallicity of the clusters and of their surrounding 
fields, especially to understand the possible formation mechanisms of these different 
kinds of SMC populations. According to Westerlund (1997), there is no reason to 
expect large differences in abundances between clusters and field stars although several 
recent studies point in the opposite direction. For example, Piatti et al. (2007a), analysing 
metallicities and ages of a sample of 42 clusters, showed that young clusters are at least 
0.3 dex more metal-rich than the population of surrounding
field stars, presumably of similar age. They interpret 
this result as evidence that most field stars are formed either from remnant gas clouds from 
star cluster formation or from disrupted clusters, in agreement with the scenario of Chandar et al. (2006). 
Nonetheless, Piatti et al. (2007c) suggested, after analysing the AMR of clusters and field stars, 
that these two populations started to undergo similar chemical enrichment histories 
the last couple of Gyrs, but their chemical evolution was clearly 
different in the period between 4 and 10 Gyr ago.

In a recent paper, Tsujimoto \& Bekki (2009) have argued that there is a dip
in the AMR of both field and cluster stars in the SMC around 7.5 Gyr ago. They 
apply chemical evolution models to suggest that this dip was caused by a major
merger of the SMC with a metal-poor, gas rich galaxy at this epoch, and find 
reasonable fits between their models and the observed AMR. 

In this paper we examine the metallicities of field stars surrounding a sample of star 
clusters of the SMC. 
As noted above, our current knowledge of the chemical evolution of this 
neighboring galaxy is very limited. In order to definitively determine the 
existence and nature of any gradient, the likelihood of a past merger, 
differences in the AMR between cluster and field stars, and other key questions
requires a substantial improvement in both data quantity and quality. The CaT
technique is a very efficient, sensitive and well calibrated metallicity index
for giant stars. Simultaneously with the cluster giants discussed in Paper I, 
we observed a large number of field giants surrounding each cluster.
This data set represents an important step in the above direction.
In section 2, we describe our field star sample, while in Section 
3 the spectroscopic observations and reduction procedures are detailed. In sections 4 
and 5, we present the radial velocities and equivalent width measurements and the metallicity 
derivation of the star fields, respectively. In section 6 we discuss the results 
obtained from the metallicities. Finally, in section 7 we summarize our main findings 
and conclusions.

\section{Field sample}

Recently, we determined metallicities and radial velocities for a sample of SMC clusters 
based on Ca II triplet spectroscopy (CaT) of red giant stars (Paper I). 
As part of Program 076.B-0553, $V$ and $I$-band pre-images of our targets were taken by 
ESO Paranal staff in August 2005. Clusters were centered on the upper (master) CCD, while the lower 
(secondary) CCD was used to observe only field stars. Target fields  were selected trying to cover as 
wide an area and radial range in the galaxy as possible in order to search for any global effects such as gradients. 
Figure 1 of Paper I shows the positions of our target sample.

The pre-images were processed within IRAF \footnote{Image Reduction and Analysis Facility, distributed 
by the National Optical Astronomy Observatories, which is operated by the Association of Universities 
for Research in Astronomy, Inc., under contract with the National Science Foundation.} and stars 
were identified and photometered using the aperture photometry routines in DAOPHOT (Stetson 1987).  
Stars were cataloged using the FIND routine in DAOPHOT and photometered with an aperture radius of 3 
pixels. The $V$ and $I$ band data were matched to form colors.

The selection of the cluster spectrocopic targets is described in detail in Paper I. In brief, they 
were chosen based on the instrumental CMD by selecting stars located along the cluster
giant branch. At the same time, we also selected as many stars as possible
on the cluster chip which also appeared to be giants falling outside the cluster
radius (after first maximizing the number of cluster stars placed on slits) in
order to explore the field star chemistry and kinematics. We similarly selected
as many field giants as possible from the secondary chip.

Field stars on the secondary chip plus the stars of the master chip which were rejected as cluster 
members according to our membership discrimination method (see Section 6 of Paper I for more 
details), are taken as the selected sample for this study.
We carefully checked the metallicities and radial velocities of the field stars taken from the master chip to make sure that
they are incompatible with the corresponding cluster values. This reaffirmed our confidence in the 
absence of star cluster contamination. The secondary chip is located far enough from the master one to make
any potential cluster contamination negligible.

The total field star sample amounts to
$\approx$ 360 stars in 15 SMC fields. Table 1 lists the cluster equatorial coordinates. 
We named each field after the corresponding 
cluster.

\section{Spectroscopic Observations and Reductions}

The spectra of the program stars were obtained during 2005 November in service mode by the VLT staff, 
using the FORS2 spectrograph in mask exchange unit (MXU) mode. Our instrumental setup 
is discussed in Paper I, which can be referred to for a more detailed description. We used 
slits that were 1'' wide and 8'' long and single exposures of 900 s were obtained with a typical 
seeing less than 1''. The spectra have a dispersion of $\sim$ 0.85 \AA/pixel (resolution of 2-3 
\AA)  with a characteristic rms scatter of $\sim$ 0.06 \AA\ and cover a range of $\sim$ 1600 \AA 
\space in the region of the CaT (8498 \AA, 8542 \AA \space and 8662 \AA). S/N values ranged from
 $\sim$ 10 to $\sim$ 70  pixel$^{-1}$. Calibration exposures, bias frames and flat-fields were also 
 taken by the VLT staff.\\

We followed the image processing detailed in Paper I. In brief, the IRAF  task {\it ccdproc} was 
used to fit and subtract the overscan region, trim the images, fix bad pixels, and flat-field each 
image. We then corrected the images for distortions, which rectifies the image of each slitlet to a 
constant range in the spatial direction and then traces the sky lines along each slitlet and puts them 
perpendicular to the dispersion direction. We used the task {\it apall} to define the sky background 
and extract the stellar spectra onto one dimension. The tasks {\it identify, refspectra} and {\it dispcor} 
were used to calculate and apply the dispersion solution for each spectrum. Finally, the spectra were
continuum-normalized by fitting a polynomial to the stellar continuum. In Figure 1 two 
examples of the final spectra in the CaT region can be observed.

\section{Radial Velocity and Equivalent Width Measurements}

Radial velocities (RVs) of our target field stars are useful for analyzing the kinematics of the SMC and 
comparing our       results with those
obtained using other SMC objects such as star clusters, carbon stars, 
etc. Although    the kinematic analysis is beyond the scope of this paper,  the program used to measure the Equivalent Width (EW) of the CaT lines requires 
knowledge of the RV to make the Doppler correction and to derive the CaT line centers.

To measure RVs of our program stars, we performed cross-correlations between their spectra and those of
32 bright Milky Way open and globular cluster template giants using the IRAF task {\it fxcor} (Tonry \& Davis 1979). 
We used the template stars of Cole et al. (2004, hereafter C04) who observed these stars with a setup very similar 
to ours. The template spectra are listed in section 4 of Paper I. {\it Fxcor} 
also makes the 
necessary correction to place the derived RV in the heliocentric reference frame. We adopted the average of 
the ensemble cross-correlation results as the heliocentric RV of a star, finding a typical standard deviation 
of $\sim$ 6 km s$^{-1}$.

Errors in centering the image in the spectrograph slit may lead to inaccuracies in determining RVs, as 
previous papers have shown (e.g., Irwin \& Tolstoy 2002). We corrected these errors following the 
procedure described in section 4 of Paper I. As shown in that section, the velocity corrections we 
have applied range from $\vert\Delta v\vert$ = 0 to 27 km s$^{-1}$ and the typical error introduced in the 
RV turns out to be $\pm$ 4.5 km s$^{-1}$. This error, added in quadrature to the one resulting from the 
cross-correlation, yields a total error of 7.5 km s$^{-1}$, which has been adopted as the typical RV 
error (random $+$ systematic) of an individual star.  

To measure EWs we have used a previously written FORTRAN program (see C04 for details). We followed the 
procedure of Armandroff \& Zinn (1988), described in detail in section 4 of Paper I, on the basis of which we 
define continuum bandpasses on both sides of each CaT line, determine the ``pseudo-continuum'' for 
each line by a linear fit to the mean value in each pair of continuum windows and calculate the 
``pseudo-equivalent width'' by fitting a function to each CaT line in relation to the pseudo-continuum.
We fit a gaussian function to each CaT line in those spectra with 
S/N $<$ 20 and fit a composite function (gaussian plus Lorentzian) to the spectra with S/N $>$ 20 
(see Paper I for justification). We 
then corrected the Gaussian-only fit for the  low S/N spectra according to equation (2) of Paper I.
 
\section{Metallicities}

The relationship between the strengths of the CaT lines and stellar abundance has been calibrated by 
several studies. In all cases, the selected CaT index uses a linear combination of the EW of two or three 
individual CaII lines to form the line strength index $\Sigma W$. Because our spectra are high  enough
quality that all three CaT lines can be measured, we adopted for $\Sigma W$ the same definition adopted 
by C04, in which all three lines are used with equal weight, namely:

\begin{equation}
\Sigma W = EW_{8498} + EW_{8542} + EW_{8662},
\end{equation}

Theoretical and empirical studies have shown that effective temperature, surface gravity and [Fe/H] 
all play a role in CaT line strengths (e.g., J\o rgensen et al. 1992; Cenarro et al. 2002). However, Armandroff \& Da Costa (1991) showed that 
there is a linear relationship between a star's absolute magnitude and $\Sigma W$ for red giants of a given 
metallicity. Following  previous authors, we define a reduced equivalent width, $W'$, to remove the 
effects of surface gravity and temperature on $\Sigma W$ via its luminosity dependence:

\begin{equation}
W' = \Sigma W + \beta (V-V_{HB}),
\end{equation}

\noindent in which the introduction of the difference between the visual magnitude of the star ($V$) and the 
cluster's horizontal branch/red clump ($V_{HB}$) also removes any dependence on distance and interstellar 
reddening. Here, as our magnitudes are uncalibrated, we use $v$ and $v_{HB}$. The $v_{HB}$ was derived from the corresponding cluster or field CMD, for stars on the master or 
secondary chip, respectively. 
In those cases where the red clump happened not to be clearly evident on the secondary chip (L\,106, L\,110 and L\,111), 
we assume that the field $v_{HB}$ is 
the same as that of the cluster located on the master chip.  As Grocholski et al. (2006) discussed in detail, the use of an inappropriate $V - V_{HB}$ can
introduce systematic errors in the derived metallicities. Specifically, C04 and Koch et al. (2006) showed that the associated error in [Fe/H] 
is on the order of $\pm$0.05 dex but it can be as large as $\pm$0.1 dex, in extreme cases. Therefore, for these three fields, we
 have added an error of $\pm$0.1 dex in quadrature 
with the one corresponding to the metallicity calculation.  

The value of $\beta$ has been investigated by 
previous authors. We prefer to adopt the value obtained by C04, i.e. $\beta$ = 0.73 $\pm$ 0.04,
because C04's instrumental setup was very similar to ours and they investigated
this parameter in depth. As discussed in detail 
in Grocholski et al. (2006), it is not necessary to make any corrections for age effects. Rutledge et al. (1997) showed that there is a 
linear relationship between the reduced EW and  metallicity on the carretta \& Gratton (1997) abundance scale for globular 
clusters of the Milky Way. C04 extended this calibration to a wider range of ages (2.5 Gyr $\leq$ age $\leq$ 13 
Gyr) and metallicities ($-2.0 \leq$ [Fe/H] $\leq -0.2$) by combining the metallicity scales of Carretta \& Gratton (1997) and 
Friel et al. (2002) for globular and open clusters, respectively. Further extensions of the calcium triplet 
calibrations are provided by Battaglia et al. (2008) (down to -2.5 dex) and by Carrera et al. (2007)  (to +0.5 dex and 0.25 Gyr).
We adopted the C04 relationship:

\begin{equation}
[Fe/H] = (-2.966 \pm 0.032) + (0.362 \pm 0.014)W',
\end{equation}

\noindent to derive the metallicities of our field star sample. We estimate that the total  metallicity 
error (random $+$ systematic) per star ranges from 0.09 to 0.35 dex, with a mean of 0.17 dex.
In Table 2 we list the information for the individual stars. Columns (1), (2)
and (3) show the identification of the star,
right ascension and declination, respectively.  Table 2 also lists
$v-v_{HB}$ in column (4), $\Sigma W$ and its error in columns (5) and
(6) and metallicity and its error in columns (7) and (8).
We considered only those stars with 50 $<$ RV $<$ 250 km s$^{-1}$ as SMC
members (Harris \& Zaritsky 2006).

We emphasize that this analysis followed the identical procedure used and 
detailed in Paper I for the cluster giants, assuring that the derived 
metallicities are completely comparable.

\section{Metallicity analysis}

The field star metallicity distribution (hereafter MD) in a galaxy is an extremely useful tool 
to investigate its overall chemical evolution. Figure 2 shows the MD of all the field stars 
in our sample as well as the (quite good) gaussian fit. We derived a mean 
metallicity of [Fe/H] = -1.00 $\pm$ 0.02 with a dispersion of 0.32 $\pm$ 0.01, in 
excellent agreement with the global mean value of -1.0 found by Carrera et al. (2008) from 
CaT spectra of a large number of field giants. Our derived mean metallicity also shows 
very good agreement with the mean value of [Fe/H] = -0.96 we found in Paper I from 
CaT spectra of star clusters.

There are previous hints in the literature about the existence of a metallicity 
gradient in the SMC. Piatti et al. (2007a,b) found that the mean metallicity values and the 
respective metallicity dispersions of their cluster sample tend to be higher for the clusters 
located within 4$^{\circ}$ from the SMC center than for those situated outside this radius. 
Recently, Carrera et al. (2008) studied $\sim$ 350 red giant stars in 13 fields distributed from 
$\sim$ 1$^{\circ}$ to $\sim$ 4$^{\circ}$ from the center, using CaT spectroscopy. They found 
a mean metallicity of [Fe/H] $\sim$ -1.0 in the {\it innermost} SMC fields, with the 
mean decreasing in the outermost regions, reaching [Fe/H] $\sim$ -1.6 at $\sim$ 
4$^{\circ}$ radius from the SMC center. They also found a relationship between this metallicity 
gradient  and the age gradient in the sense that the youngest stars, concentrated in the 
central regions, are the most metal-rich. However, in Paper I 
we found no clear evidence of any true metallicity gradient in the SMC cluster system from 
data which extend to regions futher from the center than the outermost Carrera 
fields. Cioni (2009) derived the [Fe/H] of 7653 SMC AGB stars within an area of 
20$^{\circ}$ X 20$^{\circ}$. Her results are in agreement with the lack of a metallicity gradient in
this galaxy. In addition, her mean metallicity value of -1.12 $\pm$ 0.03 is
 also in good agreement with our field value
within their dispersions.

With the aim of testing the possible existence of a metallicity gradient from our field stars, 
we fit gaussian functions to the MD of each of our fields. The resulting MDs together with their 
respective gaussian fits are shown in Figures 3, 4, 5 and 6.
Since the gaussian fits of the L\,7 (Figure 4 (d)) and L\,17 (Figure 5 (a)) fields were not satisfactory,
 we decided to use the median 
metallicity in each case. The resulting median metallicities and standard errors of the median (between brackets) are 
-1.01 (0.07) and -0.89 (0.06) for L\,7 
and L\,17, respectively.
In the L\,106 field we do not have a sample of stars large enough to fit a
gaussian to the MD (Figure 5 (d)). The 
mean value of this small sample is  [Fe/H] = -0.92 $\pm$ 0.16 (standard error of the mean). From a statistical
point of view, the L106 field sample is too small to conclude that the mean metallicity value is more
appropriate than the median value. We decided to use the mean value for the subsequent analysis, however, it is 
necessary to keep in mind that the median metallicity for this sample is -0.77.
The remaining MDs 
exhibit reasonably good single gaussian fits. In Table 3 we list field ID in column (1), the number $n$ 
of stars belonging to the field in column (2), the mean metallicity and metallicity dispersions 
with their respective errors in columns (3) and (4) and the semi-major axis $a$ (discussed below) in column (5).

In order to look into the possible existence of a metallicity gradient in the SMC, the orientation 
of the galaxy and projection effects must first be addressed. The orientation is so far poorly 
determined and the galaxy is markedly elongated along the line of sight, making the 
determination of true galactocentric distances difficult to perform. 
We then followed, as in Paper I, 
the procedure described by Piatti et al. (2007a), according to which we adopted an elliptical coordinate 
system (Figure 1, Paper I) and computed for each field the value of the semi-major axis, $a$, which 
an ellipse would have under the following conditions: (i) if it were centered on the SMC center; (ii) 
if it were aligned with the bar; (iii) had a $b$/$a$ ratio of 1/2; and (iv) if one point of the trajectory coincided with the field 
position. Then, we use the $a$ value as a surrogate for the true galactocentric distance. 

In Figure 7 we plot the metallicity vs. the semi-major axis $a$ value for our field sample (filled circles) where it is evident that no clear trend is present. 
We divided our field sample into two regions: 
{\it inner} and {\it outer} 4$^{\circ}$ from the SMC center, as was done in Piatti et al. (2007a,b)  
and in Paper I.  In Figure 7 there are nine fields in the {\it inner} group and six fields in the 
{\it outer} group. We found a mean metallicity value of [Fe/H] = -0.99 for the {\it inner} 
group and [Fe/H] = -1.02 for the {\it outer} one (standard deviations of 0.08 and 0.07, 
respectively). This result reinforces the conclusion reached in Paper I in which very similar mean metallicity 
values of -0.94 $\pm$ 0.19 (standard deviation - 15 clusters) and -1.00 $\pm$ 
0.21 (10 clusters) were 
found for clusters in
the {\it inner} and {\it outer} regions, respectively. The cluster sample of Paper I is also included in Figure 7 (open circles and triangles). 
We remind the reader that in 
Paper I we supplemented our 15 clusters observed
with FORS2 (open circles in Figure 7) with 10 additional clusters from the literature (Da Costa \& Hatzidimitriou 1998; Glatt et al. 2008; 
Gonzalez \& Wallerstein 1999) plotted as triangles. 
We showed in Paper I that the lack of a metallicity gradient 
cannot be caused by an age gradient effect since the mean ages and standard deviations are 3.1 and 1.9 
Gyr for the {\it inner} clusters and 4.4 and 3.4 Gyr for the {\it outer} ones. As 
already mentioned, Carrera et al. (2008) suggested the existence of such a gradient. In order to compare their 
results with ours, we have included in Figure 7 the fields studied by those authors 
(squares). As can be seen, their evidence for the existence of a metallicity 
gradient is completely dependent    on the two outermost fields.
There are now 7
fields and 10 clusters located in the same outer
region, whose mean metallicity is indistinguishable     from the 
global  value of [Fe/H] = -1.0. No other datapoints support such a low metallicity at any radius. In particular,
the 4 clusters and 3 field points  at even larger galactocentric distances are completely in agreement
with all of the other points except the two outliers of Carrera.
We believe that, from a statistical point of view, the absence of a 
metallicity gradient is more probable, as supported by Cioni (2009).  
In fact, the best fit line for the data shows a clear tendency to flatness. The linear fit for our field sample turned
out to be [Fe/H] = -0.006 $\pm$ 0.009 $\times a -$ 0.98 $\pm$ 0.04 with a rms = 0.08 (solid line in Figure 7).
The linear fit for the full sample (fields + clusters in Paper I) is
[Fe/H] = 0.007 $\pm$ 0.012 $\times a -$ 1.00 $\pm$ 0.05 with a rms = 0.14 (dashed line in Figure 7).
Note that the error of Carrera et al.'s two outermost fields 
are the largest of their sample. As we have previously mentioned, Piatti et al. (2007a,b) also found evidence for a metallicity gradient 
but it is important to keep in mind that the Piatti et al. values are
based on Washington photometry, which typically has error bars of about 0.2 dex,
but also includes a significant age correction for stars younger than 5 Gyr
when using their standard giant branch technique.

Carrera et al (2008) found evidence for a universal AMR. In the 
light of this, they argued that their metallicity gradient is not due to a variation of the AMR 
but to an age gradient, with  the younger stars, which are the most metal-rich, concentrated in the 
central region of the galaxy. This result reinforces previous suggestions of Piatti et al. (2007a,b) 
that the farther a cluster is from the center of the galaxy, the older and the more metal-poor it is. 
Then, if we accept that the AMR is in fact universal, the lack of a gradient of metallicity 
implies that there is not a variation of the age with the distance. It is interesting to note that 
Piatti et al (2007c) found some relatively young clusters in the outer region which 
present a new twist    
for cluster formation in the SMC and its chemical  evolution. They suggest that chemically 
enriched gas clouds can exist in the outermost portions of the galaxy. They do not discard the 
possibility that in the outer body ($a > $ 3.5$^{\circ}$) of the SMC metallicity and age gradients 
could be somewhat negligible or non-existent. Also, Cioni (2009) suggests that during an encounter of the
SMC with the LMC ($\sim$ 3 Gy ago), star formation started to take place in the 
outer parts of the galaxy altering the [Fe/H] gradient. 
Zaritsky et al. (1994) and Friedli \& Benz (1995) found that, while abundance gradients are common
in spiral galaxies, the presence of a classical bar tends to weaken the
gradient over a few dynamical timescales.  This effect is seen in the LMC which has a stellar bar and
shows no significant metallicity gradient (Olszewski et al. 1991; Geisler et al. 2003; Grocholski et al. 2006).  Thus, the
presence of a bar may explain the lack of a metallicity gradient in the SMC.
It is also necessary to bear in mind that the true distance of each star in the field from the galaxy center
is unknown. We assume the projected semimajor axis distance as the 
most appropriate coordinate system under the circumstances.
In addition, the presence of a true (i.e., 3-dimensional) radial gradient can be weakened in the transition to projected,
2-dimensional coordinates.

To compare the clusters with their surrounding fields, we have plotted
the difference between the metallicity of the field and that of the
cluster versus the $a$ value in Figure 8 and versus the cluster age in
Figure 9. The adopted cluster age can be found in Table 4 of Paper I.
Figures 8 and 9 show a clear tendency for most fields to be more
metal-poor than the corresponding cluster, independently of their
positions in the galaxy or the cluster age.  There are, however, three
exceptions to this behavior.  The fields around L\,5, L\,6, and L\,27 are all
more metal-rich than their corresponding clusters.  These three clusters
are the most metal-poor clusters in out CaT sample, and as such in
Fig.~10 we plot the metallicity difference between the fields and
clusters as a function of the cluster metallicity.  A clear trend is
seen for the metallicity difference to decrease with increasing cluster
metallicity and L 5, L 6, and L 27 are no longer outliers.  We note that
the large error associated with L 106 is due to the small number of
field stars available.

One possible explanation for this trend is if clusters showed internal
metallicity gradients and tidal disruption of the clusters stripped off
the outer, more metal-poor stars.  In this case, the resulting field
stars surrounding the clusters would have a lower metallicity.  However,
this cannot explain the presence of field stars that are more metal-rich
than their corresponding cluster, and given that the field stars have
radial velocities that are distinct from the clusters, we think that it
is unlikely that the field stars are associated with the currently
observed clusters. Also note that such metallicity gradients are not observed
in any known clusters. Our preferred explanation is that the bulk of the
field stars formed at an older epoch from most of the clusters we have
observed, with the exact epoch dependent on which AMR we adopt.  For
example, we showed in Paper I that both the clusters and the field stars
show only minor enrichment from approximately 10-3 Gyr ago, with an
average [Fe/H] $\sim -1$, followed by substantial increase in
metallicity after 3 Gyr.  Since most of the clusters we observe are
younger than $\sim$3 Gyr the observed AMR naturally explains the more
metal-poor field stars if the bulk of the field is older than $\sim$3
Gyr.  This is in agreement with the results of Sabbi et al. (2009), which
suggest that the SMC actively formed field stars over a long time
interval until about 2-3 Gyr ago.  The existence of clusters like L\,5,
L\,6, and L\,27, which are the three intermediate-age, but relatively
metal-poor clusters in our sample, is likely the result of minor mergers
with metal-poor gas clouds which diluted the ISM in the SMC only
locally, but may also be the result of a major merger that diluted the
entire SMC (e.g. Tsujimoto \& Bekki 2009).  Finally, in this scenario
the question arises as to why the field stars would be generally older
than the bulk of the clusters.  It is typically thought that most stars
are formed in clusters (e.g. Lada \& Lada 2003) and that
field stars are the result of the subsequent dissolution of a majority
of the clusters (e.g. Chandar et al. 2006; but see Bastian et al. 2009 for an
opposing view), with only the more massive clusters surviving for an
extended period of time.  Thus, it is possible that prior to about
$\sim$ 3 Gyr ago most of the clusters that formed in the SMC were of low
enough mass to be easily dispersed into the field, whereas younger
clusters were mostly massive enough to survive to the current time and
thus not contribute a significant number of young stars to the field.
This possiblity is suppoted by dynamical simulations by Bekki et al. (2004) 
which suggest that the LMC and SMC had a very close encounter
$\sim$ 4 Gyr ago that would have resulted in an increased rate of
massive cluster formation in the Magellanic Clouds.  Note, however, that
the results of Besla et al. (2007) have brought into
question whether or not the Magellanic Clouds have been interacting for
an extended period of time.

Of course, information about the age of the fields is needed to perform
a more reliable analysis of the chemical evolution history of field stars and to compare it
with the evolution of the cluster system.

\section{Summary and Conclusions}

We used VLT + FORS2 to obtain near-infrared spectra of $\sim$ 360 giant stars distributed in 
15 SMC fields, covering the spectral range that includes the three \ion{Ca}{2} triplet lines. 
From these spectra we derived individual star radial velocity and
metallicity, applying the CaT technique 
(Cole et al. 2004; Grocholski et al. 2006; Parisi et al. 2009)
in exactly the same manner as followed in Paper I for the targeted cluster.
A mean error of 0.17 dex is achieved. The following summarizes the results 
of the analysis of these metallicities:

i) By fitting a Gaussian function to the whole sample, we found a mean value of [Fe/H] 
= -1.00 $\pm$ 0.02 with a dispersion of 0.32 $\pm$ 0.01. This metallicity is in good 
agreement with the mean value of [Fe/H] = -1.0 found by Carrera et al. (2008) for 
field stars and [Fe/H] = 
-0.96 reported in Paper I for star clusters.

(ii) We also fit gaussian functions to the metallicity distribution of each field in order to derive 
their mean metallicity and dispersion. There is practically no deviation from the value 
of [Fe/H] = -1.0 of the individual mean metallicities, which range from [Fe/H] = -0.88 to 
[Fe/H] = -1.11. The dispersions vary from 0.13 to 0.46. 

(iii) By dividing our sample into {\it inner} and {\it outer} 4$^{\circ}$ from the SMC center, 
we found a mean metallicity (standard deviation) of [Fe/H] = -0.99 (0.08) and [Fe/H] = 
-1.02 (0.07) for the {\it inner} and {\it outer} regions, respectively. This result suggests that 
there is not a metallicity gradient in the SMC, in agreement with the work of Cioni (2009) 
and with Paper I and contrary to the trend suggested by Carrera et al. (2008). Carrera et al. (2008) found evidence for a universal 
AMR and suggested that the metallicity gradient they derived is due to the presence of 
an age gradient in the galaxy. If we assume a universal AMR, the fact that our metallicities 
do not show any tendency to vary according to the distance from the SMC center suggests 
that there is not an age variation either. This is consistent with the possible scenario 
presented by Piatti et al. (2007c), who derived the age of some outer clusters and found 
that they are indeed young objects.
The lack of a metallicity gradient in our data can also be explained by the
presence of a classical bar which tends to weaken the gradient (Zaritsky et al. 1994: Friedli \& Benz 1995).
This effect has also been seen in the LMC (Olszewski et al. 1991; Geisler et al. 2003; Grocholski et al. 2006).

(iv) From the comparison between the metallicity of the star fields and that of the clusters 
they surround, it is evident that there exists a tendency for the fields to be more 
metal-poor than the clusters, independently of the age of the cluster and of its position 
in the galaxy. 
We argue that this is due to        the clusters 
covering a range of both ages and metallicities but mainly younger and more
metal-rich, while the field stars may have dated from an older 
epoch lasting many Gyr in which the metallicity was almost uniform
 and more metal-poor.
Of course, information about the age of the fields is needed to perform
a more reliable analysis of the chemical evolution history of field stars and to compare it
with the evolution of the cluster system.
 
\acknowledgments

The paper was improved by a number of useful comments from the referee.
This work is based on observations collected at the European Southern Observatory,
Chile, under program number 076.B-0533. We would like to thank the 
Paranal Science Operations Staff. 
D.G. gratefully acknowledges support from the Chilean
Centro de Astrof\'\i sica FONDAP No. 15010003
and the Chilean Centro de Excelencia en Astrof\'\i sica
y Tecnolog{\'\i}as Afines (CATA). 
M.C.P. and J.J.C. gratefully acknowledge 
financial support from the Argentinian 
institutions CONICET, FONCYT and SECYT (Universidad Nacional de C\'ordoba).

{\it Facilities:} \facility{VLT: Antu (FORS2)}.

\begin{figure}
\plotone{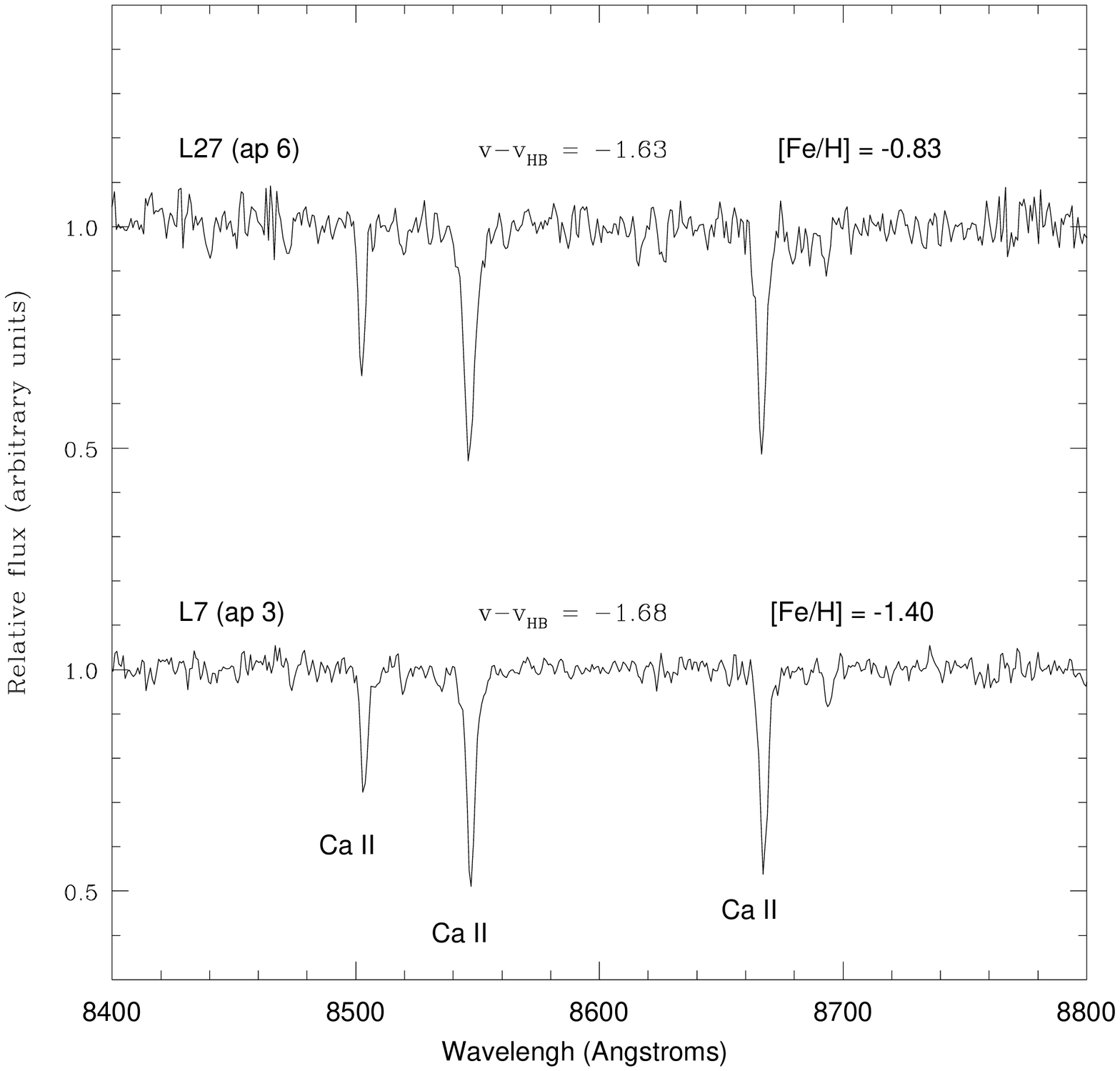}
\caption{Sample continuum-normalised spectra of   RGB stars in two fields of our sample. The three
CaT lines have been marked on the plot as well as the corresponding $v-v_{HB}$
values and metallicities. These 2 stars have very similar $T_{effs}$ and log g
values. Thus, the difference in Ca II line strength illustrates their
substantial metallicity difference. }
\end{figure}

\begin{figure}
\plotone{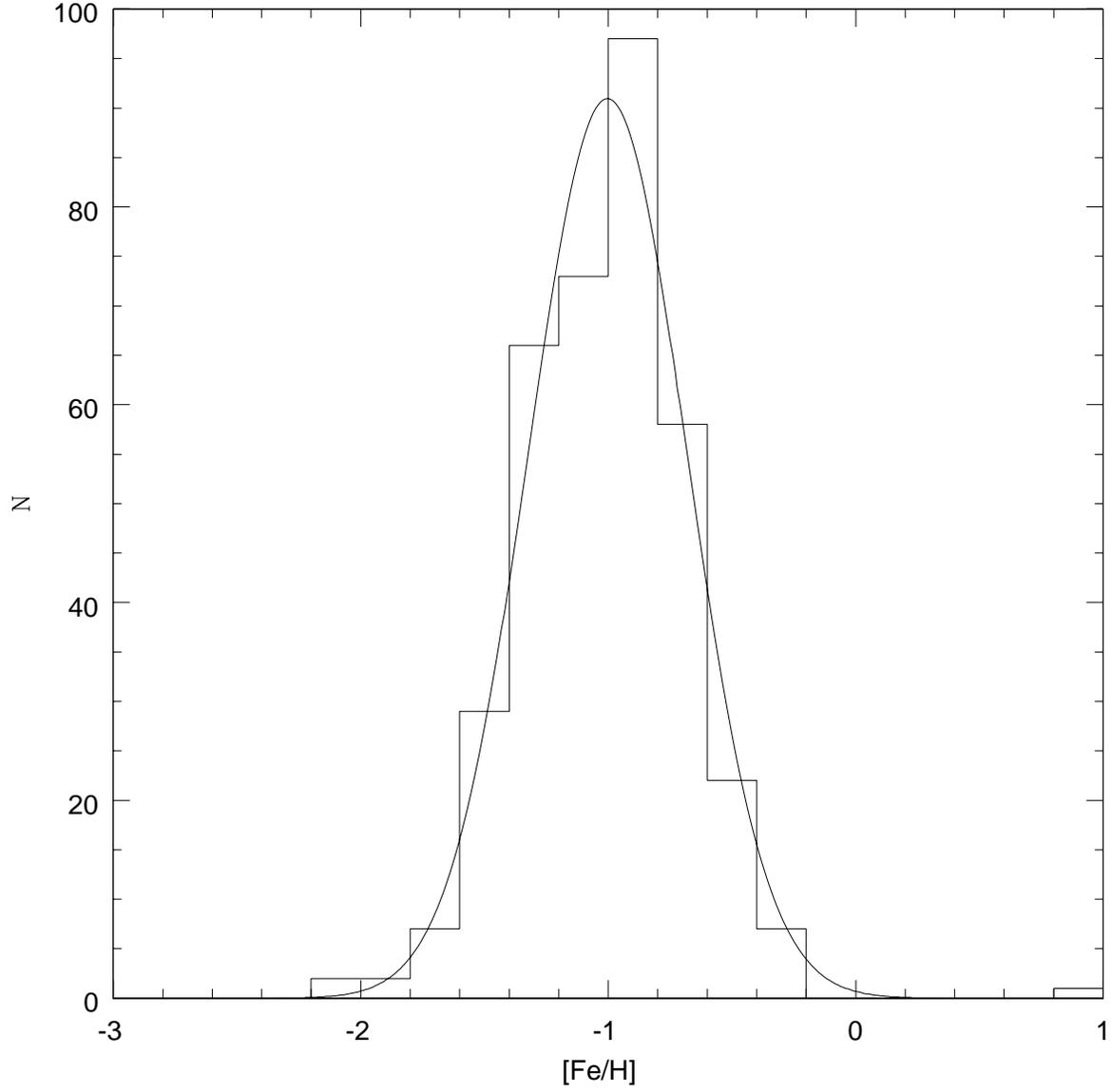}
\caption{Metallicity distribution of all field stars of our sample. The corresponding gaussian
fit is shown by the solid line.}
\end{figure}

\begin{figure}
\plotone{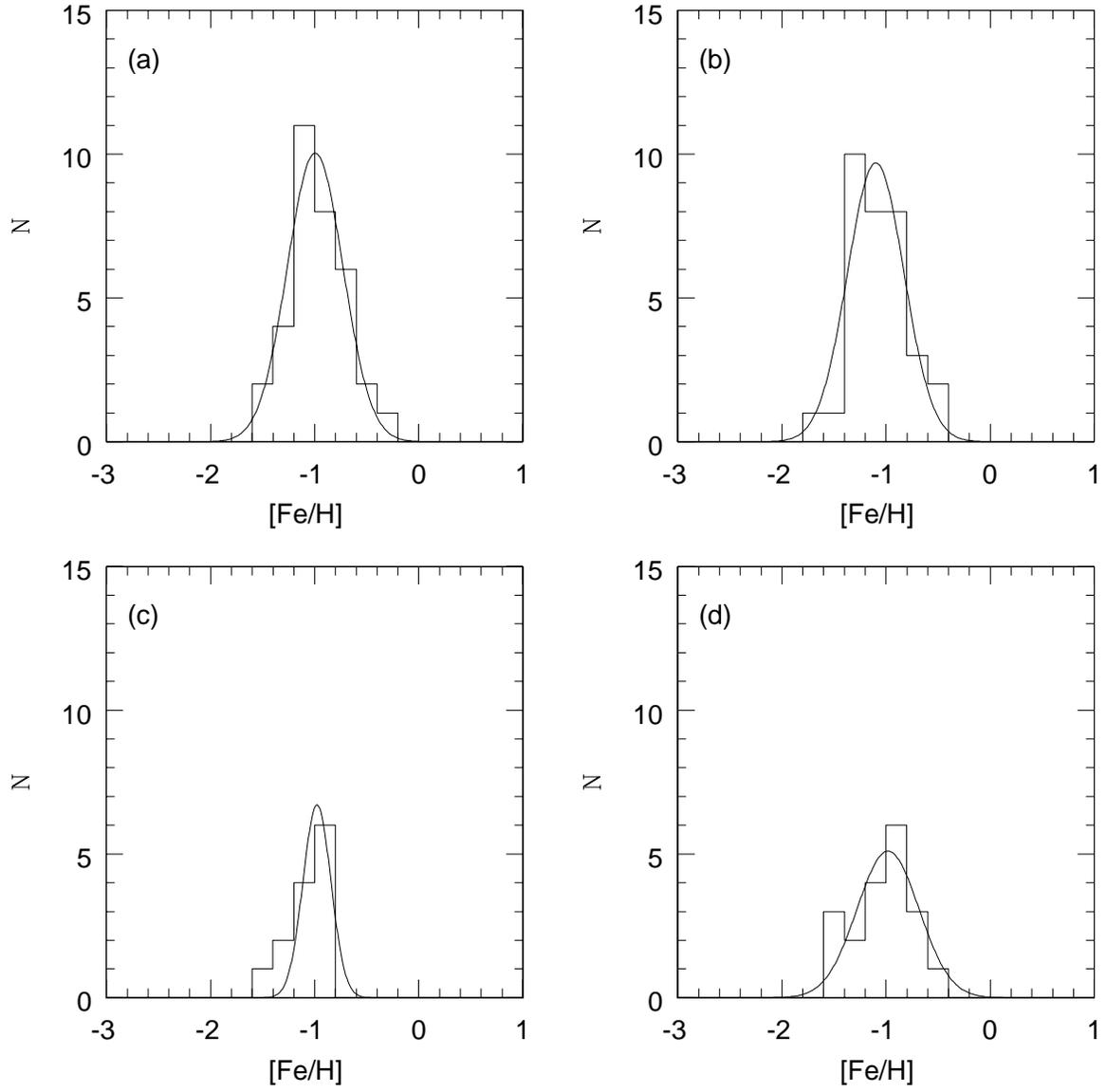}
\caption{Metallicity distributions of the field stars surrounding the clusters: (a) BS\,121, (b) HW\,47,
(c) HW\,84 and (d) HW\,86. The solid curves show the corresponding gaussian fit.
}
\end{figure}

\begin{figure}
\plotone{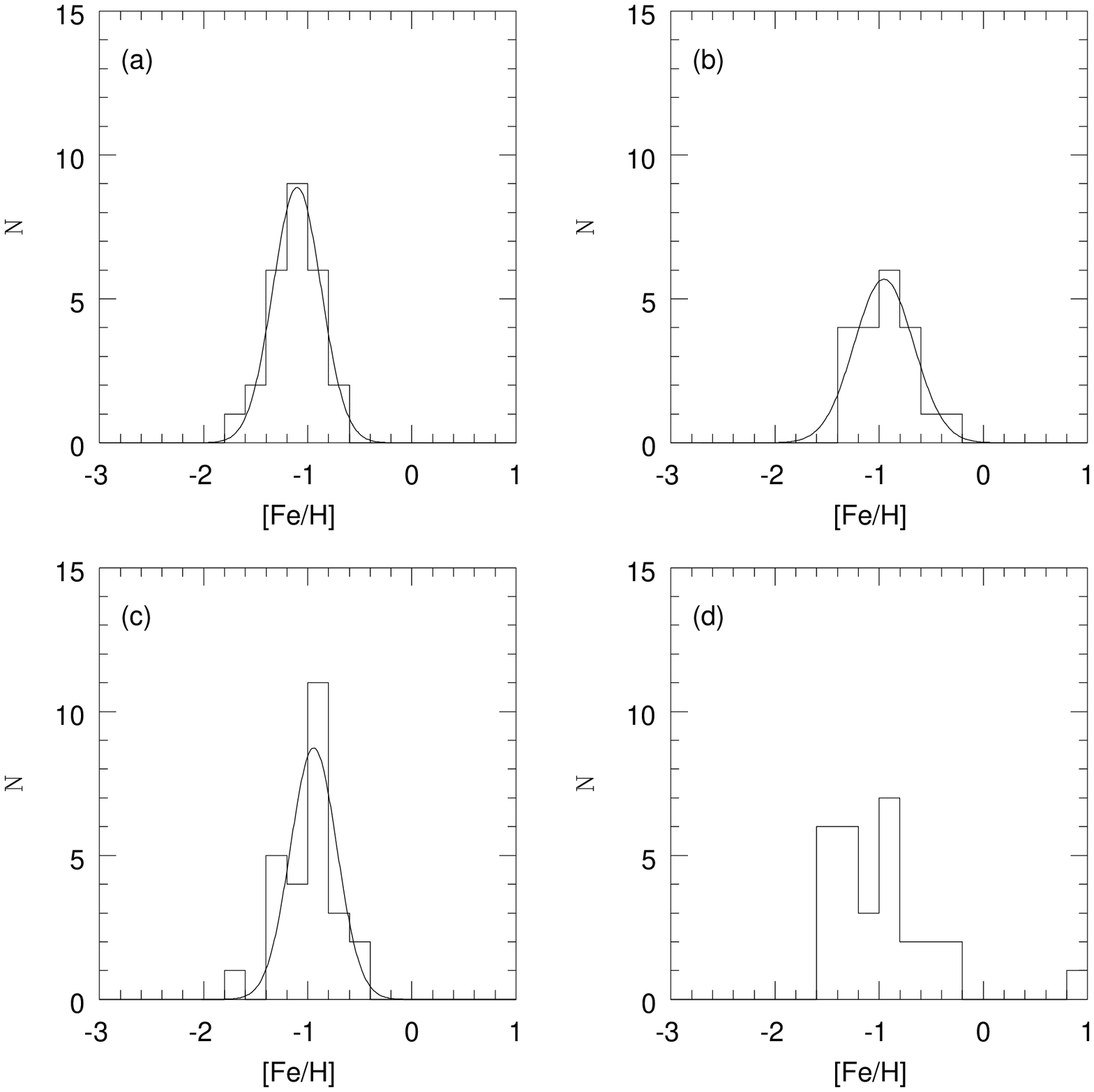}
\caption{Same as in Figure 3 for the field stars surrounding the clusters: (a) L\,4, (b) L\,5, (c) L\,6
and (d) L\,7.
 }
\end{figure}

\begin{figure}
\plotone{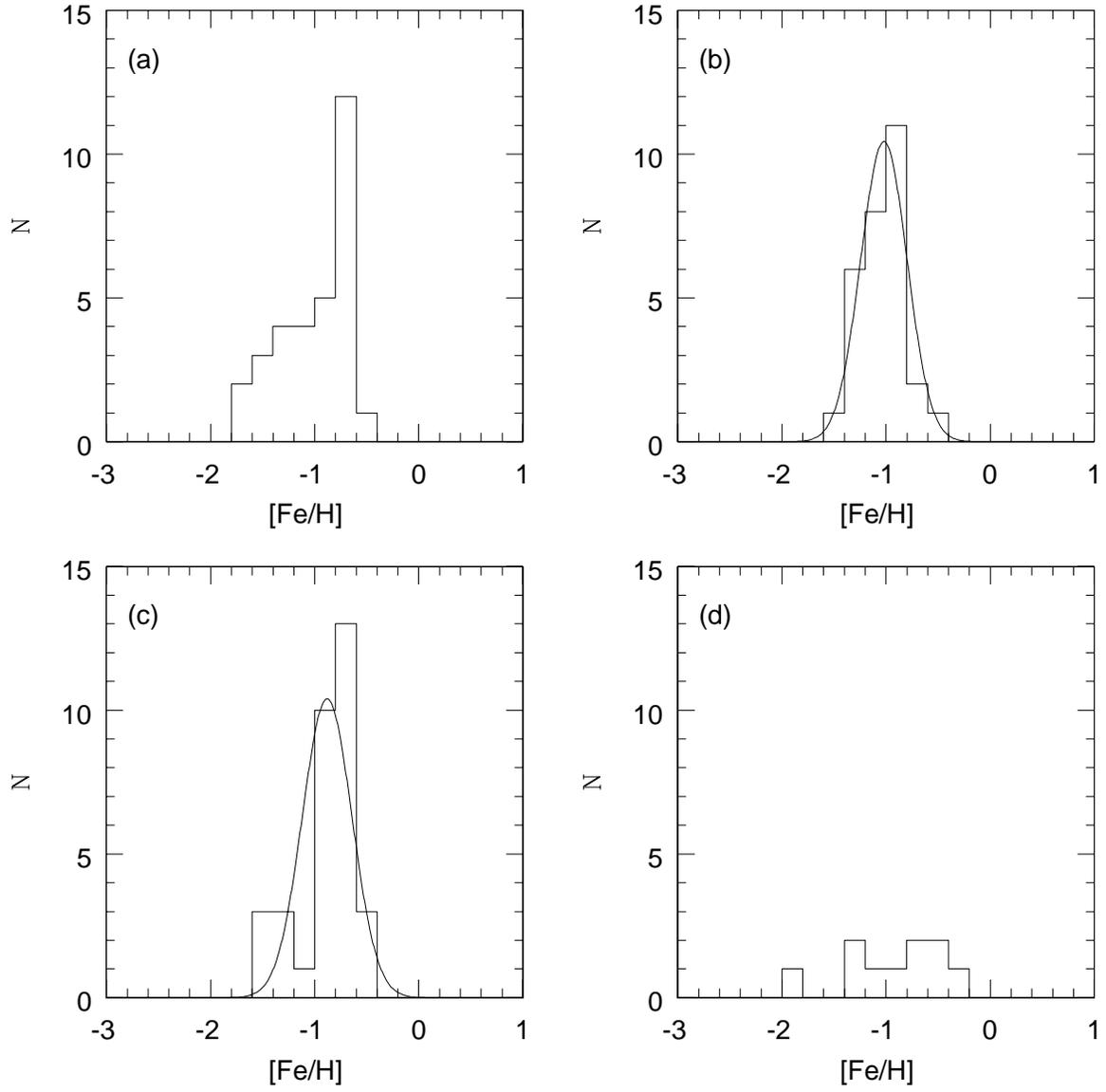}
\caption{Same as in Figure 3 for field stars surrounding the clusters: (a) L\,17, (b) L\,19, (c) L\,27
and (d) L\,106.
}
\end{figure}

\begin{figure}
\plotone{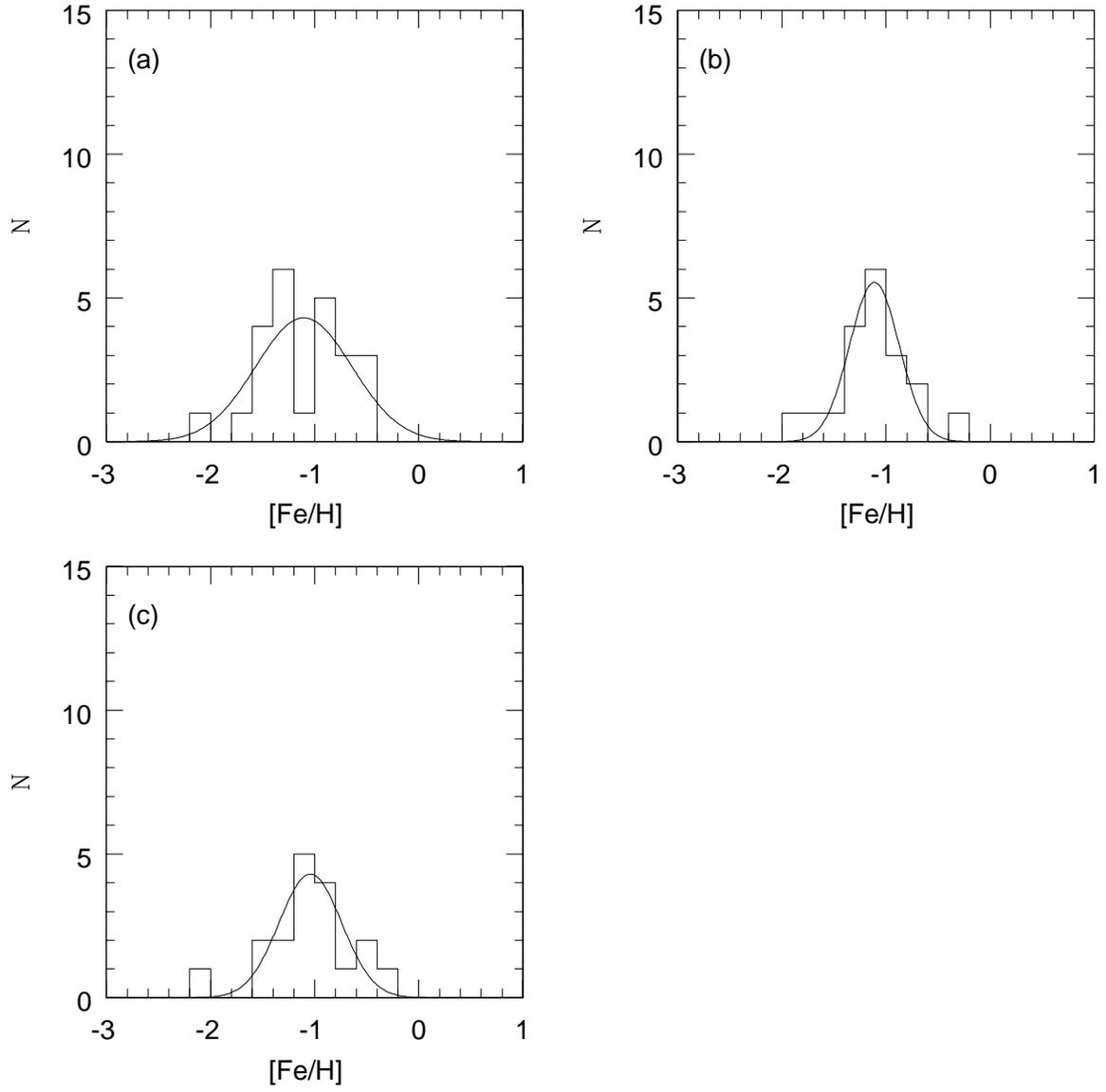}
\caption{Same as in Figure 3 for the field stars surrounding the clusters: (a) L\,108, (b) L\,110 and (c) L\,111.
}
\end{figure}

\begin{figure}
\plotone{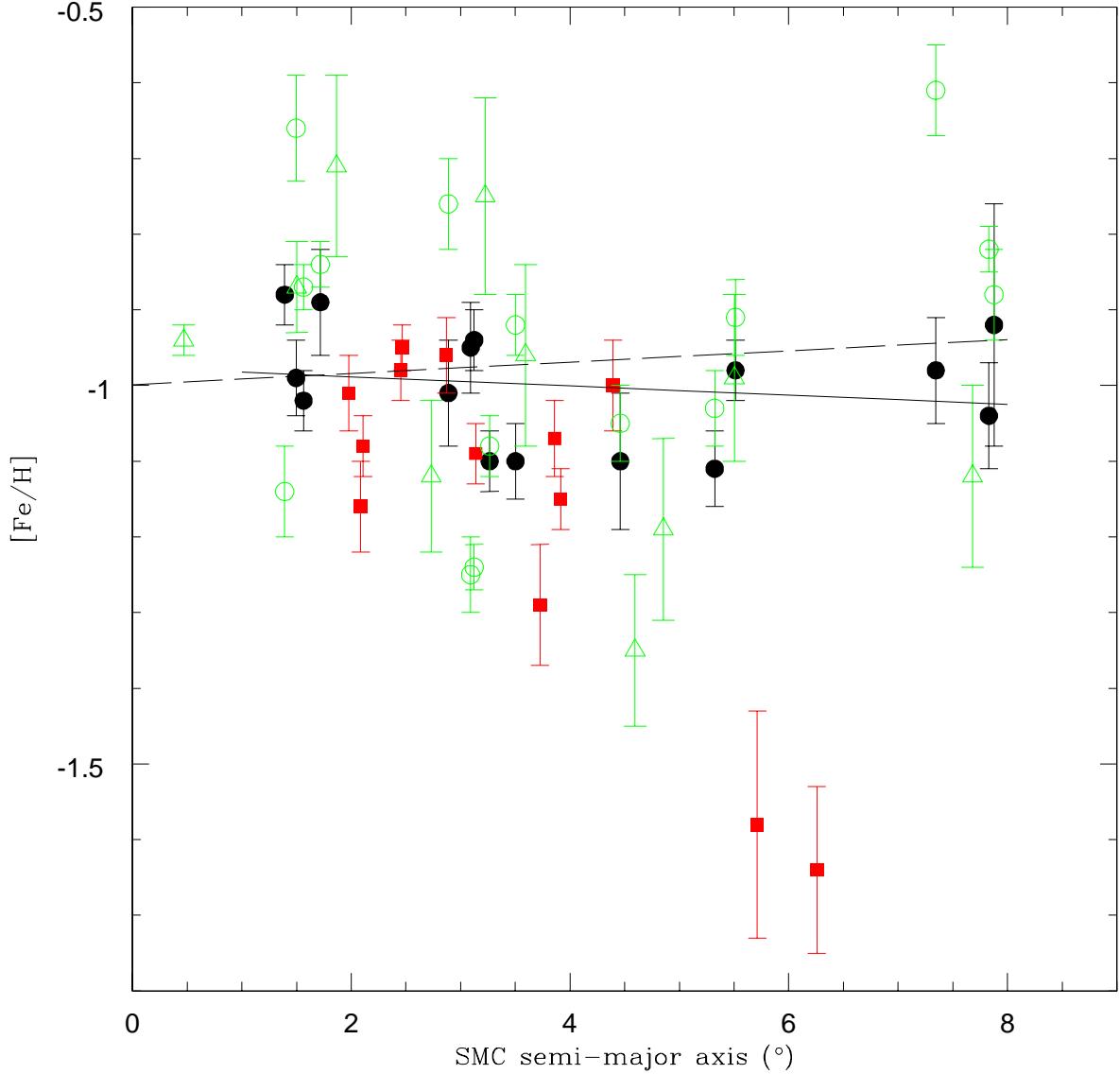}
\caption{Mean metallicity vs. semi-major axis $a$ for the fields of our sample (filled circles).
Open circles represent our CaT cluster sample studied in Paper I, and empty triangles correspond 
to the additional cluster sample also included in Paper I. Fields of Carrera et al. (2008) 
are represented by squares. Error bars correspond to the standard error of the mean except for the
extended cluster sample which errors are the ones quoted in the correspondig paper 
(Da Costa \& Hatzidimitriou 1998; Glatt et al. 2008;
Gonzalez \& Wallerstein 1999).
Solid line is the linear fit for our field sample while the dashed one correspond to the linear fit
of the complete sample (fields plus clusters in Paper I).
The colour figure is available electronically.
}
\end{figure}

\begin{figure}
\plotone{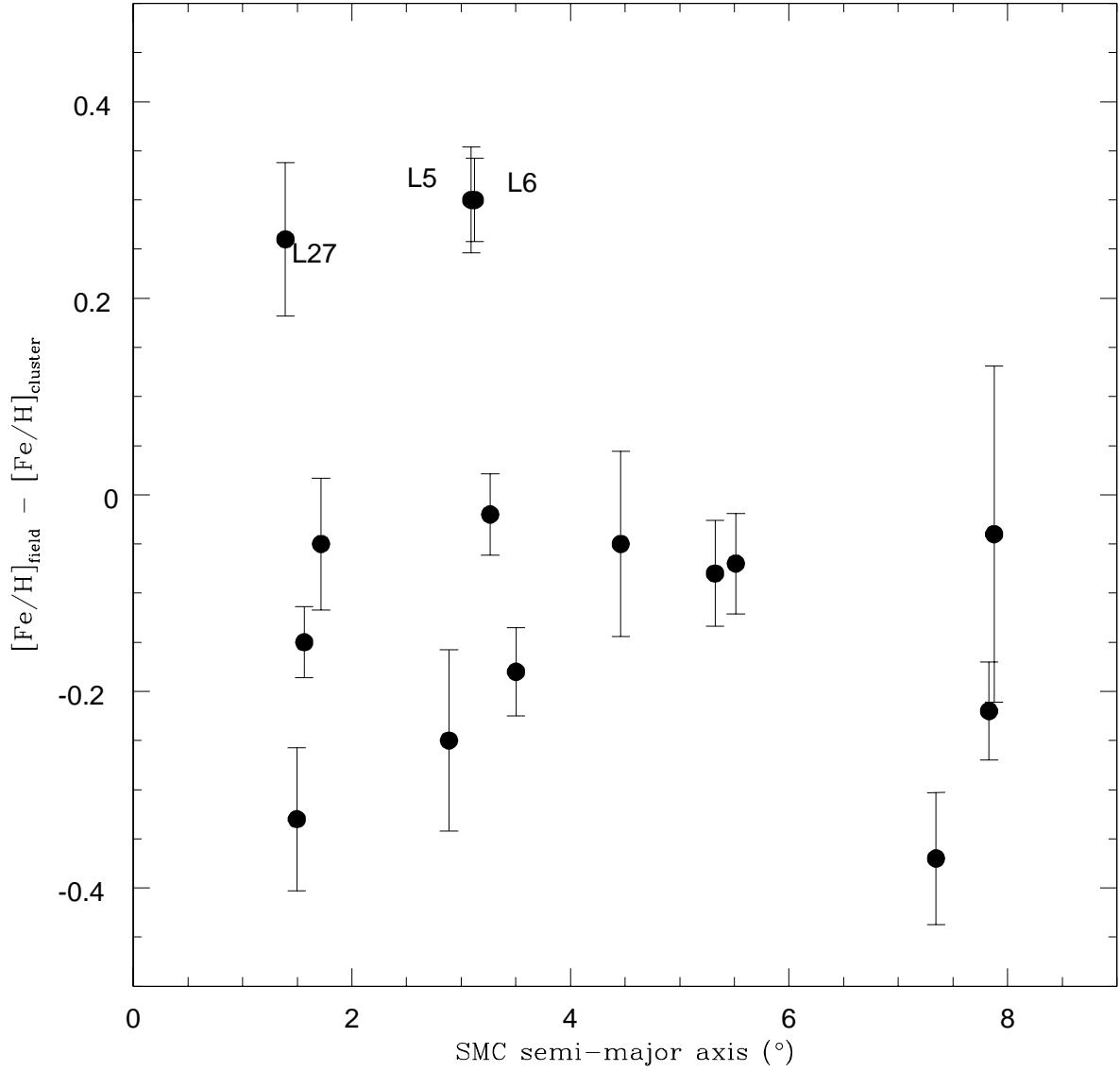}
\caption{Difference between the metallicity of fields and that of the corresponding 
clusters versus the semi-major axis $a$.
}
\end{figure}

\begin{figure}
\plotone{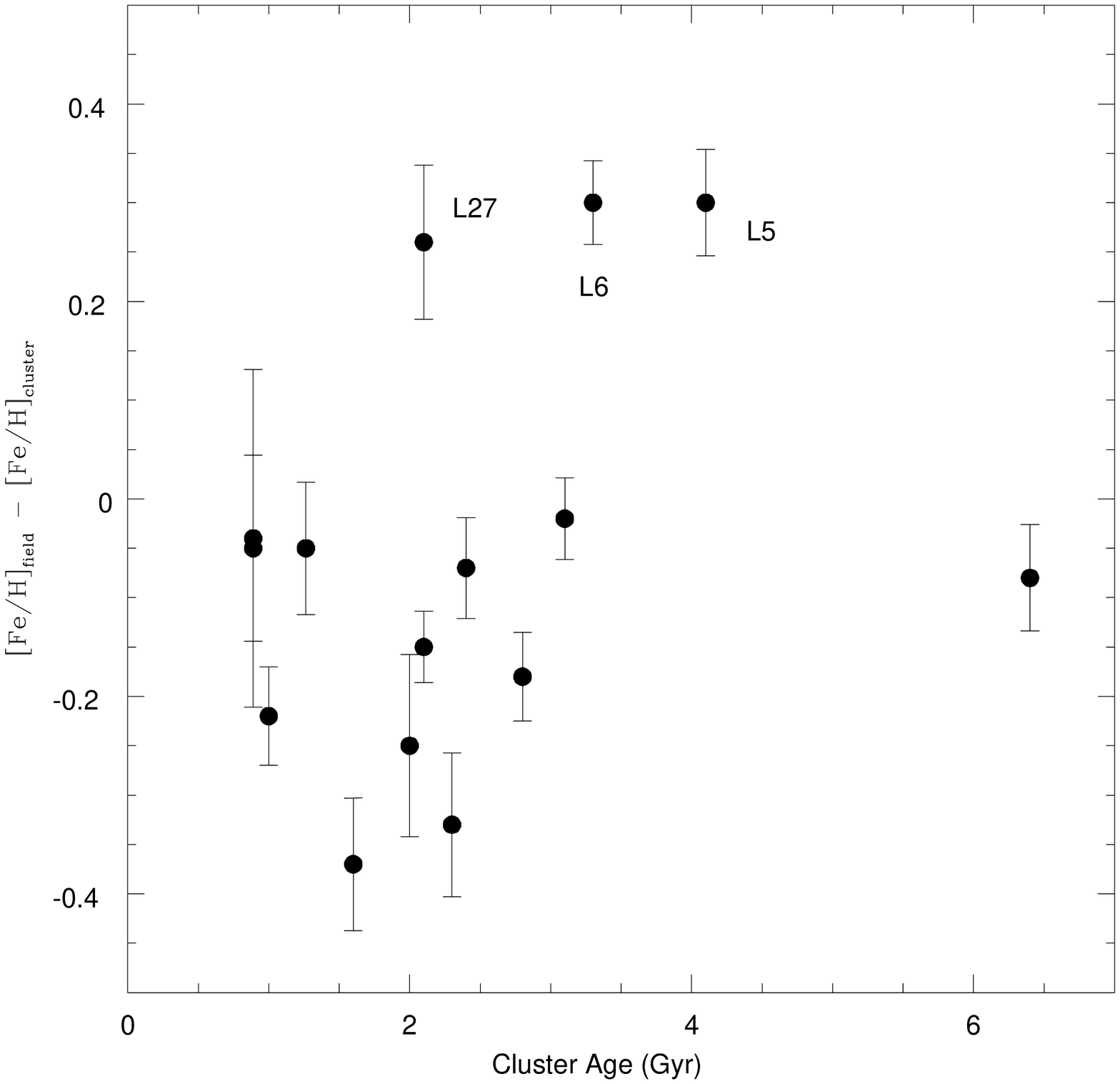}
\caption{Difference between the metallicity of fields and that of the corresponding clusters 
versus the cluster age.}
\end{figure}

\begin{figure}
\plotone{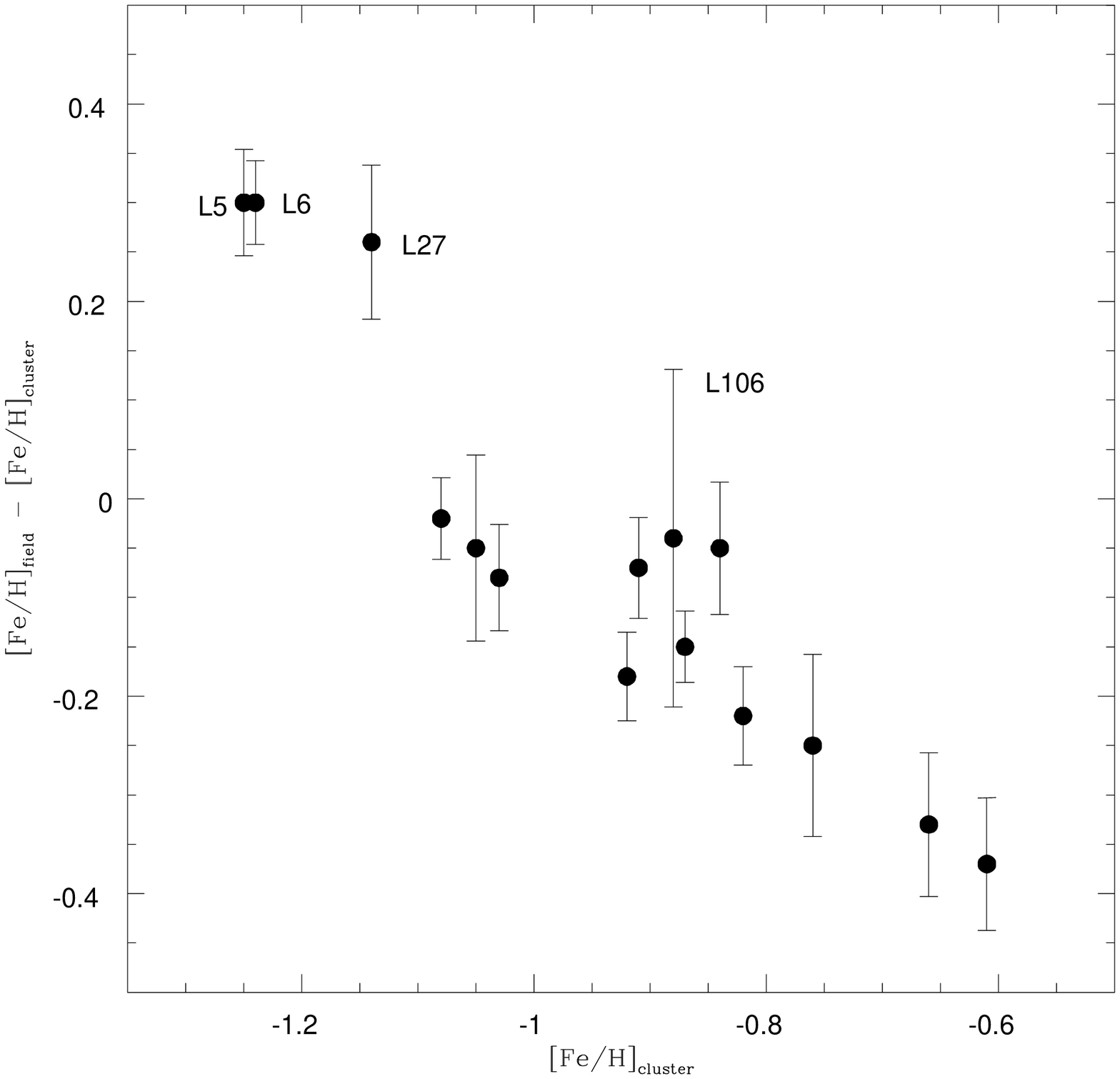}
\caption{Difference between the metallicity of fields and that of the corresponding clusters 
versus the cluster metallicity.}
\end{figure}

\begin{deluxetable}{lcc}
\tablewidth{0pt}
\tablecaption{SMC Clusters}
\tablehead{
\colhead{Cluster}                & \colhead{RA (J2000.0)}  &
\colhead{Dec (J2000.0)}                                    \\
                                 & \colhead{($h$ $m$ $s$)} & 
\colhead{($^{\circ}$ $'$ $''$)}                            }
\startdata
BS\,121                             & 01 04 22 & -72 50 52 \\
HW\,47                                               & 01 04 04 & -74 37 09 \\
HW\,84                                               & 01 41 28 & -71 09 58 \\
HW\,86                                               & 01 42 22 & -74 10 24 \\
L\,4                               & 00 21 27 & -73 44 55 \\
L\,5                                     & 00 22 40 & -75 04 29 \\
L\,6                               & 00 23 04 & -73 40 11 \\
L\,7                               & 00 24 43 & -73 45 18 \\
L\,17                              & 00 35 42 & -73 35 51 \\
L\,19                                   & 00 37 42 & -73 54 30 \\
L\,27                           & 00 41 24 & -72 53 27 \\
L\,72                            & 01 03 53 & -72 49 34 \\
L\,106                                   & 01 30 38 & -76 03 16 \\
L\,108                                               & 01 31 32 & -71 57 10 \\
L\,110                                    & 01 34 26 & -72 52 28 \\
L\,111                          & 01 35 00 & -75 33 24 \\
\enddata
\end{deluxetable}

\begin{deluxetable}{lccccccc}
\tablewidth{0pt}
\tablecaption{Position and Measured Values for Field Stars}
\tablehead{
\colhead{ID}                    & \colhead{RA (J2000.0)}        &
\colhead{Dec (J2000.0)}         & \colhead{v-v$_{HB}$}          &
\colhead{$\Sigma W$}             & \colhead{$\sigma_{\Sigma W}$}&
\colhead{[Fe/H]}                 & \colhead{$\sigma_{[Fe/H]}$} \\
                                & \colhead{($h$ $m$ $s$)}       &
\colhead{($^{\circ}$ $'$ $''$)} & \colhead{(mag)}               &
\colhead{(\AA)}                 & \colhead{(\AA)}               &
                                &                               }
\startdata
BS\,121M-1 & 01 04 06.51 & -72 51 09.61 &   -0.40 &  5.89 &  0.32 & -0.939 & 0.150 \\
BS\,121M-3 & 01 04 11.76 & -72 51 13.40 &   -0.21 &  5.27 &  0.42 & -1.114 & 0.177 \\
BS\,121M-6 & 01 04 23.74 & -72 50 50.82 &   -0.04 &  4.01 &  0.38 & -1.525 & 0.158 \\
BS\,121M-8 & 01 04 22.44 & -72 51 29.66 &   -1.18 &  5.59 &  0.11 & -1.253 & 0.098 \\
BS\,121M-9 & 01 04 30.69 & -72 50 57.06 &   -1.25 &  6.29 &  0.12 & -1.018 & 0.105 \\
\enddata
\tablecomments{Table 2 is published in its entirety in the
electronic edition of the {\it Astronomical Journal}.  A portion is
shown here for guidance regarding its form and content.}

\end{deluxetable}

\begin{deluxetable}{lcccccc}
\tablewidth{0pt}
\tablecaption{SMC fields results}
\tablehead{
\colhead{ID}           & \colhead{n}                    &
\colhead{$\bar{[Fe/H]}$}    & \colhead{$\sigma_{\bar{[Fe/H]}}$} &
\colhead{$a$}      }
\startdata
BS\,121 & 34 & -0.99 $\pm$ 0.02   & 0.27 $\pm$ 0.02  & 1.496 \\
HW\,47  & 33 & -1.10 $\pm$ 0.02   & 0.27 $\pm$ 0.03  & 3.502\\
HW\,84  & 13 & -0.98 $\pm$ 0.01   & 0.13 $\pm$ 0.02  & 5.513\\
HW\,86  & 19 & -0.98 $\pm$ 0.03   & 0.29 $\pm$ 0.03  & 7.345\\
L\,4    & 26 & -1.10 $\pm$ 0.01   & 0.23 $\pm$ 0.005 & 3.265\\
L\,5    & 20 & -0.95 $\pm$ 0.02   & 0.29 $\pm$ 0.02  & 3.092\\
L\,6    & 26 & -0.94 $\pm$ 0.03   & 0.22 $\pm$ 0.03  & 3.124\\
L\,7    & 29 & -1.01 $\pm$ 0.07   &    \nodata       & 2.888\\
L\,17   & 31 & -0.89 $\pm$ 0.06   &    \nodata       & 1.718 \\
L\,19   & 29 & -1.02 $\pm$ 0.02   & 0.22 $\pm$ 0.02  & 1.564\\
L\,27   & 33 & -0.88 $\pm$ 0.05   & 0.24 $\pm$ 0.04  & 1.392   \\
L\,106  & 10 & -0.92 $\pm$ 0.16   &     \nodata      & 7.877 \\
L\,108  & 24 & -1.10 $\pm$ 0.08   & 0.46 $\pm$ 0.08  & 4.460\\
L\,110  & 19 & -1.11 $\pm$ 0.02   & 0.24 $\pm$ 0.02  & 5.323\\
L\,111  & 18 & -1.04 $\pm$ 0.04   & 0.30 $\pm$ 0.04  & 7.830\\
\tableline
\enddata
\end{deluxetable}

\end{document}